\documentclass[aps,prl,twocolumn,superscriptaddress]{revtex4}
\usepackage{amsfonts,amssymb,amsmath}
\usepackage{color}
\usepackage{graphicx}
\usepackage{dcolumn}
\usepackage{bm}

\begin{document}

\title{Unbiased sampling of globular lattice proteins in three dimensions}

\author{Jesper Lykke Jacobsen}
\affiliation{LPTMS, UMR CNRS 8626, Universit\'e Paris Sud, 91405 Orsay, France}
\affiliation{Service de Physique Th\'eorique, URA CNRS 2306, CEA Saclay, 91191 Gif sur Yvette, France}

\date{\today}

\begin{abstract}

  We present a Monte Carlo method that allows efficient and unbiased
  sampling of Hamiltonian walks on a cubic lattice. Such walks are
  self-avoiding and visit each lattice site exactly once. They are
  often used as simple models of globular proteins, upon adding
  suitable local interactions. Our algorithm can easily be equipped
  with such interactions, but we study here mainly the flexible
  homopolymer case where each conformation is generated with uniform
  probability. We argue that the algorithm is ergodic and has
  dynamical exponent $z=0$. We then use it to study polymers of size
  up to $64^3 = 262\,144$ monomers. Results are presented for the
  effective interaction between end points, and the interaction with
  the boundaries of the system.

\end{abstract}


\maketitle

Self-avoiding random walks of various types are ubiquitous in Nature
and often serve as a first step in modeling polymeric systems
\cite{deGennes}. In a good solvent, polymers assume spatially extended
conformations which are well described by the usual Self-Avoiding Walk
(SAW) model. In this model, each monomer is assigned a fugacity
$\beta$, and the SAW is recovered at the smallest value $\beta_{\rm
  c}$ of $\beta$ so that the mean length of the walk diverges. This is
a critical point, meaning that observables exhibit power law behaviors
that are independent of the precise microscopic model. Therefore, the
latter is conveniently defined on a hypercubic lattice in $d$
dimensions.

By contrast, proteins in their native globular state assume compact
confirmations that occupy space as densely as possible. This suggests
modeling such biopolymers as Hamiltonian Walks (HW); by definition
these are self-avoiding walks which are constrained to visit each of
the lattice sites exactly once. Upon adding further local
interactions, the HW model has been proposed as a description of
protein melting \cite{Flory} (with bending rigidity), or of protein
folding \cite{Dill} (with suitable interactions among amino acids).
The simplest example of the latter is the so-called HP model in which
only two types of amino acids (Hydrophobic or Polar) are taken into
account.

In $d=2$ dimensions an amazing number of exact results on
self-avoiding walk models are known, due in particular to the
application of Conformal Field Theory (see \cite{review} for a recent
review). For instance, both the HW model and the Flory model of
protein melting have been solved \cite{KondevCFT}.  But in the
physically more relevant case of $d=3$, such exact results are not
available, and one must resort to approximate techniques, or to numerics,
to extract information about the model at hand.

Much work on such lattice proteins is based on exact enumeration
studies in which all possible conformations on a small cube of size
$N=L \times L \times L$ are generated. Unfortunately, despite of the HW
constraint, the number of conformations is so large that even the case
$L=4$ is out of reach \cite{Pande}. Needless to say, such small sizes
are very sensitive to the peculiarities of the lattice model (e.g.,
the choice of lattice), and cannot be considered representative of
real proteins.

Therefore, an efficient and unbiased importance sampling scheme of the
Monte Carlo (MC) type would be most welcome to study larger $L$ and
make precise statements about the thermodynamic limit $L\to\infty$.
For the usual SAW case, this is furnished by the so-called pivot
algorithm \cite{pivot}. In each move, a randomly chosen lattice
symmetry is applied to the part of the walk subsequent to a randomly
chosen pivot point. The move is accepted if the resulting walk is
self-avoiding. For $N\to\infty$, the acceptance ratio goes to zero,
but the few accepted moves suffice to decorrelate the system in time
$\sim N$. But in the HW limit, the acceptance ratio becomes
identically zero, and so the pivot algorithm is useless in this case.
Alternative growth-type algorithms are not guaranteed to yield
unbiased sampling \cite{Zhang}.

In this Letter we propose an MC algorithm which is specifically
adapted to HW, and we test it in details for the $d=3$ flexible
homopolymer case.  We show that it satisfies detailed balance and give
evidence for ergodicity; each conformation is therefore generated with
uniform probability. In contrast to the pivot algorithm, each move is
accepted with probability one. One move takes time $\sim N$, and $N$
moves are sufficient to completely decorrelate the system. Using our
algorithm, we generate conformations of size $N=262\,144$ monomers with
ample statistics, and give a number of results on their geometrical
properties.

{\bf Algorithm.} Let ${\cal C}$ be a Hamiltonian walk on the cubic
lattice.  Choose one of its two end points randomly and uniformly;
call it ${\bf x}$.  Note that site ${\bf x}$ is adjacent to one
occupied and five empty links.  Choose one of the five empty links
with uniform probability; call it $({\bf x}{\bf y})$. Then among the
occupied links adjacent to ${\bf y}$ there is exactly one which forms
part of a loop in ${\cal C} \cup ({\bf x}{\bf y})$; call it $({\bf
  y}{\bf z})$. [Note that if ${\bf y}$ is the other end point of
${\cal C}$ there is one occupied link adjacent to ${\bf y}$; otherwise
there are two.] Change ${\cal C}$ by adding $({\bf x}{\bf y})$ and
removing $({\bf y}{\bf z})$.

\begin{figure}
\includegraphics[width=7cm,angle=0]{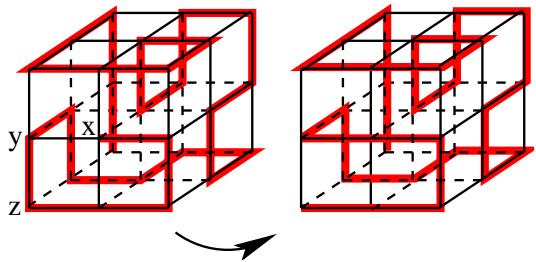}
\caption{Move between two of the $2\,480\,304$ possible Hamiltonian walks
on a $3 \times 3 \times 3$ cubic lattice.}
\label{fig:move}
\end{figure}

The MC move just described is illustrated in Fig.~\ref{fig:move}. It
was studied in two dimensions in \cite{Oberdorf}. In what follows we
choose the system to be confined to an $L^3$ cube with free (i.e., non
periodic) boundary conditions. A small modification is then needed if
${\bf y}$ is outside the cube. In that case, the move consists in
leaving ${\cal C}$ unchanged.

{\bf Unbiased sampling.} The above move clearly satisfies detailed
balance. To prove that it is also ergodic it suffices to show that any
${\cal C}$ can be transformed into a fixed reference configuration
${\cal C}_0$ in a finite number of moves. Let ${\cal C}_0$ be one of
the two configurations with a maximal number of occupied horizontal
links. One may approach ${\cal C}_0$ by choosing always ${\bf y}$ so
that $({\bf x}{\bf y})$ is horizontal.  Unfortunately, $({\bf y}{\bf
  z})$ may also turn out to be horizontal, so positive progress is not
made in every step. Moreover, if both end points end up on the left or
right side of the cube, and are both adjacent to an occupied
horizontal link, one will actually have to remove a horizontal link to
proceed. Due to these complications we cannot prove ergodicity for
arbitrary size $L$.

However, we can argue in favor of ergodicity by testing it for small
lattices.  Using exact enumeration techniques, we find that there are
$2\,480\,304$ HW on an $L=3$ cube \cite{enumeration}, and our MC
algorithm generates all of them. We also find that there are
$3\,918\,744$ Hamiltonian Circuits (HC), or closed walks, on a $3
\times 3 \times 4$ parallelepiped \cite{enumeration}. A HC can be
formed from a HW whose end points are nearest neighbors on the
lattice, by adding the missing link. In this way we have checked that
all $3\,918\,744$ HC can be generated as well.  Similar checks
have been made on smaller parallelepipeds.

Assuming that ergodicity holds in general, we conclude that the MC
process converges towards the equilibrium distribution, i.e., that all
Hamiltonian walks are sampled with uniform probability.

{\bf Performance.} Identifying the correct edge $({\bf y}{\bf z})$ to
be removed necessitates tracing out the loop formed when adding ${\bf
  x}{\bf y}$, and so one move takes a time $\sim N$. But note that
when the end point moves from ${\bf x}$ to ${\bf z}$, the part of the
chain $[{\bf x}{\bf z}{\bf y}]$ is reversed and becomes $[{\bf z}{\bf
  x}{\bf y}]$. Thus, if one were to use the algorithm to study a
heteropolymer problem, the sequence of amino acids on that part would
have to be reversed as well. So in that respect, and in terms of chain
connectivity, the move is certainly more non-local than it looks at
first sight.

Define now the autocorrelation function $G(t)$ as the link overlap
function with respect to some reference initial configuration. The
time $t$ is measured in units of MC moves per site, and averages are
done over $100$ independent runs. We have measured $G(t)$ for system
sizes $L=4,8,16,32$ and found that it decays exponentially towards an
$L$-dependent constant, which can be determined numerically by
averaging $G(t)$ over the last 100 time units in a sufficiently long
simulation. Subtracting this constant, and rescaling, gives a
normalized autocorrelation function $\tilde{G}(t)$ satisfying
$\tilde{G}(0)=1$ and $\tilde{G}(t) \to 0$ as $t \to \infty$. Defining
now the autocorrelation time $\tau$ by $\tilde{G}(t+\tau)=\frac12
\tilde{G}(t)$, we estimate $\tau$ by averaging over the first $4\tau$
of the simulation. For the system sizes mentioned this gives
$\tau=1.33$, $1.27$, $1.20$, and $1.15$.  This fits nicely as $\tau
\sim L^z$ with dynamical exponent $z=-0.024 \pm 0.001$. Of course we
cannot have $z<0$, since one MC move changes only a finite number
(two) of links. We therefore conclude that $z \approx 0$.

{\bf Simulations.} We have generated conformations on lattices up to
size $L=64$ (i.e., $262\,144$ monomers) with ample statistics. In all
cases, the first $t=100$ MC moves per site were discarded. The various
averages and histograms were then made over $T$ MC moves per site,
with measurements being made after each individual MC move. We have
taken $T=10^6$ for $L \le 20$, $T=10^5$ for $20 < L \le 30$,
$T=10^4$ for $30 < L \le 40$, $T=10^3$ for $40 < L \le 60$, and
$T=10^2$ for $L > 60$. Data were generated for both even and
odd $L$ in order to detect any parity effects.

{\bf Asymptotic isotropy.} We have measured the histogram for the
end-to-end distance vector ${\bf x}_{12} = \pm({\bf x}_1-{\bf x}_2)$;
note that the sign is immaterial because of the indistinguishability
of the end points. Not all values of ${\bf x}_{12}$ are possible,
since for $N$ even (resp.\ odd) the end points must belong to
different (resp.\ identical) sublattices by an easy parity argument.

We first checked for $L=4$ that the histogram is isotropic on the
microscopic level. Thus, for each ${\bf x}_{12}$ respecting the parity
constraint, the 6, 8, 12 or 24 vectors related to it by lattice
symmetries are found to be generated with the same frequency (up to
counting statistics fluctuations).

Still, vectors ${\bf x}_{12}$ of equal length but unrelated by lattice
symmetries need not have the same probability for finite $L$. Define
for instance the probability ratio $\rho = P\big({\bf x}_{12} =
(0,0,3)\big) / P\big({\bf x}_{12} = (1,2,2)\big)$. We find
$\rho=1.642$ for $L=4$, and $\rho=1.160$ for $L=16$. A power law fit
yields however $\rho \to 1.00 \pm 0.01$ for $L\to\infty$, so isotropy
is recovered in the thermodynamical limit. It is important for a
realistic modeling that the large $L$ behavior does not depend on
irrelevant microscopic details.

\begin{figure}
\includegraphics[width=7cm,angle=270]{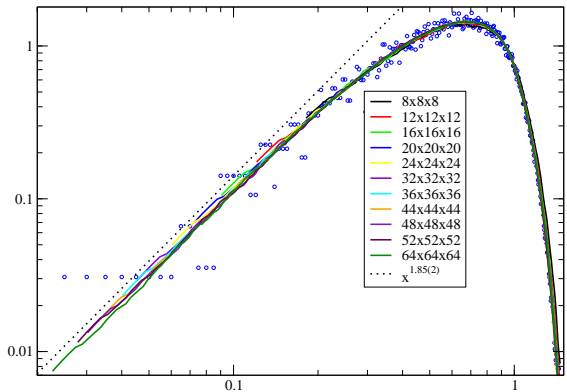}
\vskip-1.0cm
\caption{Normalized probability distribution of $x_{\rm ee} \equiv
  x_{12}/L$, where $x_{12}$ is the end-to-end distance.}
\label{fig:end2end}
\end{figure}

{\bf End-to-end distance.} We now construct a histogram for the
end-to-end distance $x_{12} = |{\bf x}_{12}|$ by summing over all
possible directions of ${\bf x}_{12}$. The discrete histogram is then
smoothed in a two-step process. First, for each value of
$x_{12}^{(0)}$ in steps of $0.1$, we regroup the counts with
$|x_{12}-x_{12}^{(0)}| < 0.5$. Second, we construct a running average
over 20 subsequent $x_{12}^{(0)}$ (i.e., two lattice spacings). Once
properly normalized, the end result is the probability distribution
(universal scaling function) of the rescaled variable $x_{\rm ee}
\equiv x_{12}/L$. This is shown in Fig.~\ref{fig:end2end}, where (as
in subsequent figures) the small circles are the $L=20$ data points
after the {\em first} step of the smoothing procedure. The data
collapse is strikingly good over the entire $x_{\rm ee}$ range.
Before the onset of finite-size effects, we have a power law behavior
\begin{equation}
 P(x_{\rm ee}) \sim x_{\rm ee}^{1.85 \pm 0.02}
 \mbox{ for } x_{\rm ee} \ll 1 \,.
 \label{Pxee}
\end{equation}
If the end points were independent we would have $P(x_{\rm ee}) \sim
x_{\rm ee}^{d-1}$ by simple geometry. The result (\ref{Pxee}) then
indicates a weak effective attraction between end points.

{\bf Conformational exponents.} For a HW of length $N$, the radius of
gyration $R_{\rm g} \sim L$, and the exponent $\nu$ appearing in
$R_{\rm g} \sim N^\nu$ is trivially $\nu=\frac{1}{d}=\frac13$.

The exponent $\gamma$ describes the ratio between the number of HW
($\sim \mu^N N^{\gamma-1}$) and HC ($\sim \mu^N N^{-\nu d}$) passing
through a fixed point. This leads to
\begin{equation}
 P(x_{12}=1) \sim L^{-\gamma d} \,, \qquad
 \gamma = 0.96 \pm 0.01 \,.
 \label{Px12}
\end{equation}
Standard scaling \cite{deGennes} gives $\gamma=1-\Delta_1$ and
$P(x_{\rm ee}) \sim x_{\rm ee}^{2 - 2\Delta_1}$, and comparing
(\ref{Pxee})--(\ref{Px12}) yields then our final
estimate $\Delta_1 = 0.06 \pm 0.02$.

{\bf Surface effects.} We next examine the
interaction of the HW with the surfaces of the system. Although
globular proteins would be more realistically modeled within a
spherical domain (where we expect our algorithm to be ergodic as
well), we maintain here the geometry of a cube in order to have
several types of surface sites.

\begin{figure}
\includegraphics[width=7cm,angle=270]{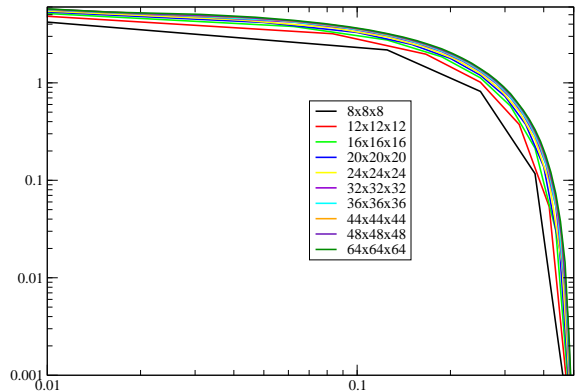}
\vskip-1.0cm
\caption{Normalized probability distribution of $x_{\rm S} \equiv
  d_{\rm S}/L$, where $d_{\rm S}$ is the distance to the nearest side
  of the cube.}
\label{fig:side}
\end{figure}

\begin{figure}
\vskip-0.3cm
\includegraphics[width=7cm,angle=270]{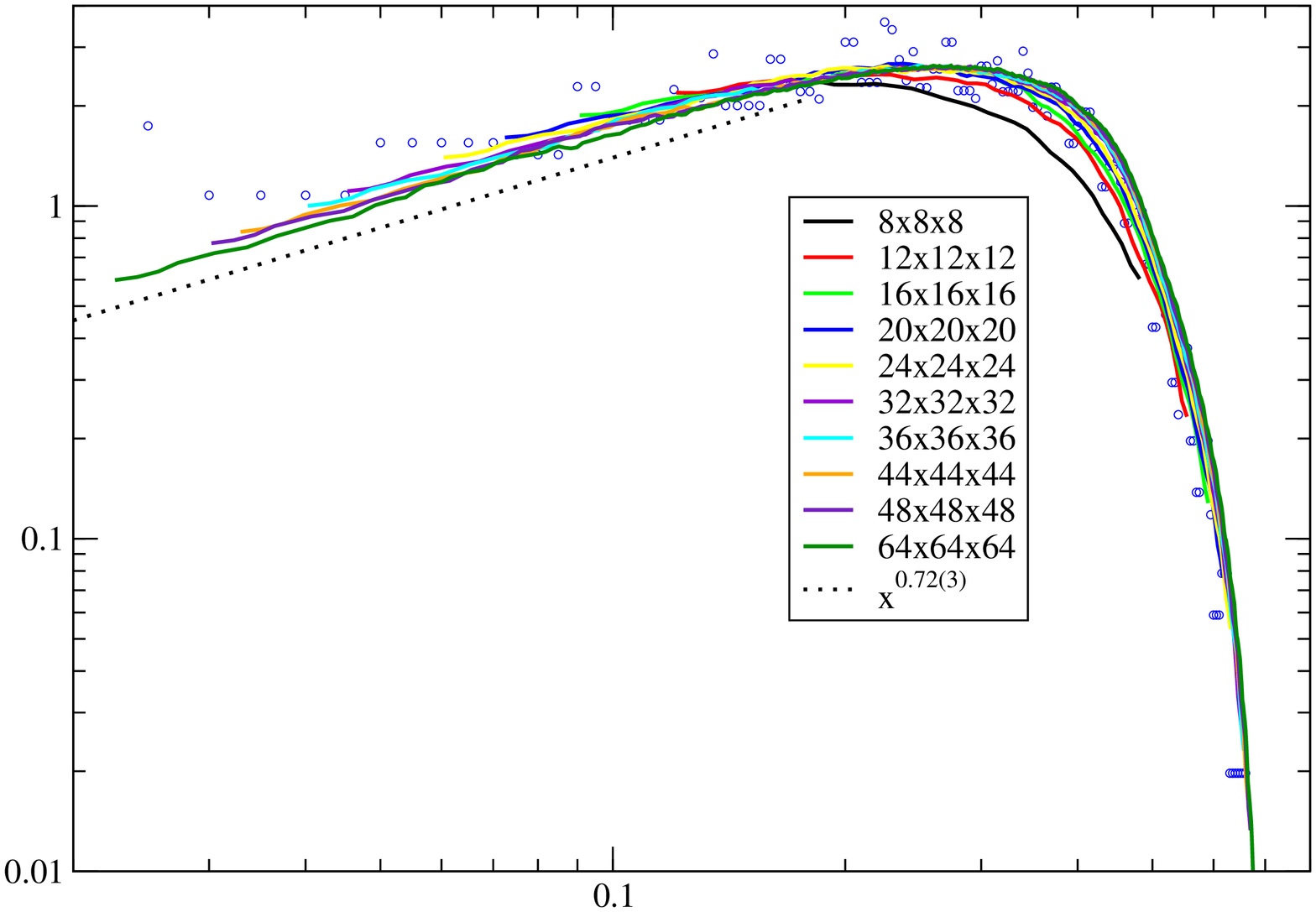}
\vskip-1.0cm
\caption{Normalized probability distribution of $x_{\rm E} \equiv
  d_{\rm E}/L$, where $d_{\rm E}$ is the distance to the nearest edge
  of the cube.}
\label{fig:edge}
\end{figure}

\begin{figure}
\vskip-1.0cm
\includegraphics[width=7cm,angle=270]{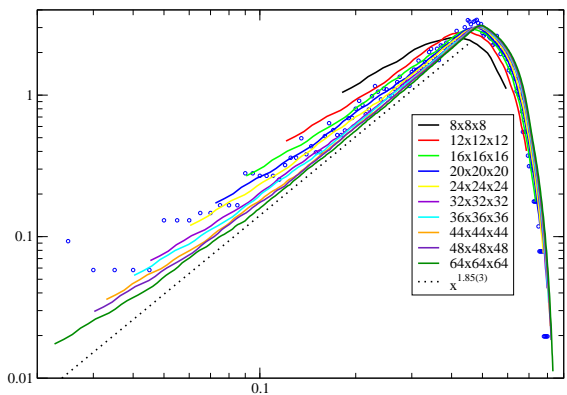}
\vskip-1.0cm
\caption{Normalized probability distribution of $x_{\rm C} \equiv
  d_{\rm C}/L$, where $d_{\rm C}$ is the distance to the nearest corner
  of the cube.}
\label{fig:corner}
\end{figure}

For a given end point, let $d_{\rm S}$, $d_{\rm E}$, and $d_{\rm C}$
be respectively its distance to the nearest side, edge, and corner of
the cube. For each of these we construct the normalized probability
distribution as above. [Since $d_{\rm S}$ is an integer, we do not
apply the smoothing procedure in this case.] The results, shown in
Figs.~\ref{fig:side}--\ref{fig:corner}, in all cases lead to very
satisfactory data collapses. Power-law fits analogous to (\ref{Pxee})
give $P(x_{\rm E}) \sim x_{\rm E}^{0.72\pm0.03}$ and
$P(x_{\rm C}) \sim x_{\rm C}^{1.85\pm0.03}$; the geometric exponents
for a non-interacting system would be $1$ and $2$ respectively.

The behavior of $x_{\rm S} \equiv d_{\rm S}/L$ in Fig.~\ref{fig:side}
would seem in favor of $P(x_{\rm S}) \to p_{\rm S}^0$ as $x_{\rm S}
\to 0$, with a {\em finite} value $p_{\rm S}^0$. However, using only
the data with $d_{\rm S}=0$, we estimate $P(d_{\rm S}=0) \sim L^{-0.94
  \pm 0.02}$, implying that $P(x_{\rm S}) \sim x_{\rm S}^{0.06 \pm
  0.02}$.

\begin{figure}
\vskip-1.0cm
\includegraphics[width=7cm,angle=270]{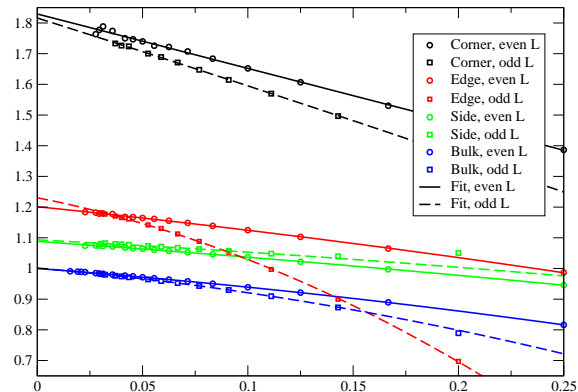}
\vskip-1.0cm
\caption{Normalized probabilities for an end point being a corner site
  ($p_{\rm C}$), an edge site ($p_{\rm E}$), a side site ($p_{\rm
    S}$), or a bulk site ($p_{\rm B}$), as functions of $1/L$.}
\label{fig:surface}
\end{figure}

To check in more detail the affinities of the end points to the boundary,
we divide the lattice sites in four classes:
\begin{equation}
 \begin{array}{lll}
   \mbox{C:} & 8 \mbox{ corner sites }     &
   \mbox{with } d_{\rm C}=0 \\
   \mbox{E:} & 12(L-2) \mbox{ edge sites}  &
   \mbox{with } d_{\rm E}=0, \mbox{ but } d_{\rm C}>0 \\
   \mbox{S:} & 6(L-2)^2 \mbox{ side sites} &
   \mbox{with } d_{\rm S}=0, \mbox{ but } d_{\rm E}>0 \\
   \mbox{B:} & (L-2)^3 \mbox{ bulk sites}  &
   \mbox{with } d_{\rm S}>0 \\
 \end{array} \nonumber
\end{equation}
Equivalently, the coordination number (number of links adjacent to a
site) is 3 for C-sites, 4 for E-sites, 5 for S-sites, and 6 for
B-sites.

We normalize the probability $p_i$ of an end point belonging to a
given class $i={\rm C,E,S,B}$ by the ratio of $i$-type sites on the
lattice. These probabilities are shown as functions of $1/L$ in
Fig.~\ref{fig:surface}. Note that for even (resp.\ odd) $L$, a HW can
link any given corner site to 3 (resp.\ 7) other corners.
Therefore, we plot $p_{\rm C}$ for even $L$ and $\frac12 p_{\rm C}$ for
odd $L$. With this convention, parity effects are seen to
vanish as $L \to \infty$, within error bars, and we find
$p_{\rm C} = 1.82 \pm 0.02$,
$p_{\rm E} = 1.21 \pm 0.01$,
$p_{\rm S} = 1.089 \pm 0.002$, and
$p_{\rm B} = 1.0000 \pm 0.0001$.
Thus, the higher the coordination number of a site, the more do the end
points tend to avoid it (in our normalization). This is remarkable,
since naively one might have expected the opposite to occur.

One may similarly study the normalized probabilities $p_{ij}$ that
both end points belong to prescribed classes $i$ and $j$. From this
we form the matrix of cross correlations $A$ with entries
$A_{ij} = \frac{p_{ij}}{p_i p_j}-1$. Note that if the two end points
were statistically independent, all $A_{ij}$ would be zero. For $L=19$
we find
\begin{equation}
 A = \left[ \begin{array}{rrrr}
 -0.100 & -0.020 & -0.006 &  0.004 \\
        & -0.020 & -0.005 &  0.003 \\
        &        & -0.004 &  0.002 \\
        &        &        & -0.001 \\
 \end{array} \right]
\end{equation}
Although the entries are numerically small, there is a clear qualitative
effect. Namely, if one end point is at a site of small coordination number,
the other end point will try to compensate by being at a site of high
coordination number. In particular, it is ten percent less likely to find
both end points in corners than would have been expected from the data for
just one end point.

{\bf Conclusion.} We have presented an algorithm that allows for
efficient and unbiased sampling of three-dimensional lattice protein
conformations (Hamiltonian walks). We applied it
to the flexible homopolymer case, where each conformation has the same
energy, and extracted a number of results on the correlation between
end points and their interaction with the boundaries of the system.
The most salient features are 1) the weak entropic attraction between end
points, 2) the attraction of end points towards the surface, edge sites
in particular, 3) their preference for sites with low coordination number,
and 4) the vanishing of parity effects for $L\to\infty$.

We hope to study later topological issues, such as the knot
formation probability.

It is a straightforward extension of our algorithm to associate an
energy to each state and implement a standard Metropolis dynamics.
This would allow to study models with bending rigidity,
amino acid interactions, etc.

A slightly modified algorithm in which end points are allowed to grow
or retract can be applied to models of non-compact conformations.  In
the standard SAW case, it decorrelates a polymer of length $N$ in time
$\sim N$, not only globally (as does the pivot algorithm \cite{pivot})
but also locally. We shall report more on this elsewhere.

\smallskip

{\bf Acknowledgments.} This work was supported through the European
Community Network ENRAGE (grant MRTN-CT-2004-005616) and by the Agence
Nationale de la Recherche (grant ANR-06-BLAN-0124-03).

\end{document}